**Preliminary Analysis of Construction Work Zone on Roadways in Florida by Crash Severity**

**Tatiana Deslouches**
Undergraduate Student
Department of Electronic Engineering Technology
Florida A&M University, Tallahassee, Fl 32307
Email: tatiana1.deslouches@famu.edu

**Doreen Kobelo Regalado, Ph.D.**  https://orcid.org/0009-0008-9020-6227
Associate Professor
Department of Construction/Civil Engineering Technology
Florida A&M University, Tallahassee, Fl 32307
Email: doreen.kobelo@famu.edu

**Mohamed Khalafalla, Ph.D., MBA, PMP**
Assistant Professor
Department of Construction/Civil Engineering Technology
Florida A&M University, Tallahassee, Fl 32307
Email: mohamed.ahmed@famu.edu

**Tejal Mulay, Ph.D.**
Assistant Professor
Department of Electronic Engineering Technology
Florida A&M University, Tallahassee, Fl 32307
Email: tejal.mulay@famu.edu

Word Count: 4,396 words + 4 table (250 words per table) = 5,396 words

**ABSTRACT**
Construction zones are inherently hazardous, posing significant risks to construction workers and motorists. Despite existing safety measures, construction zones continue to witness fatalities and serious injuries, imposing economic burdens. Addressing these issues requires understanding root causes and implementing preventive strategies centered around the 4Es (Engineering, Education, Enforcement, Emergency Response) and 4Is (Information Intelligence, Innovation, Insight into communities, Investment, and Policies). Proper safety management, integrating these strategic initiatives, aims to reduce and potentially eliminate fatalities and serious injuries in work zones. In Florida, road construction work zone fatalities and serious injuries remain a critical concern, especially in urban counties. Despite a $12 billion infrastructure investment in 2022, Florida ranks eighth nationally for fatal work zone crashes involving commercial motor vehicles (CMVs). Analysis from 2019 to 2023 shows an average of 71 fatalities and 309 serious injuries annually in Florida's work zones, reflecting a persistent safety challenge. High-risk counties include Orange, Broward, Duval, Hillsborough, Pasco, Miami-Dade, Seminole, Manatee, Palm Beach, and Lake. This study presents a preliminary analysis of work zone crashes in Broward, Duval, Hillsborough, and Orange counties. A multilogit model assessed attributes contributing to fatalities and serious injuries, such as crash type, weather and light conditions, work zone type, type of shoulder, presence of workers, and law enforcement. Results indicate significant contributing factors, highlighting opportunities to use machine learning for alerting drivers and construction managers, ultimately enhancing safety protocols and reducing fatalities.

**Keywords:** Construction work zones, crash severity, road safety, Florida.

**INTRODUCTION**

Construction zones are hazardous environments that expose construction workers and motorists to high-risk zones. Despite current safety measures, fatalities and serious injuries in construction zones remain a significant global concern, impacting lives and imposing economic costs [1, 2]. The key strategies used to reduce these incidents include the 4Es (Engineering, Education, Enforcement, and Emergency Response) and the 4Is (Information Intelligence, Innovation, Insight into communities, and Investment and Policies). Proper safety management requires integrating these strategies to reduce and potentially eliminate fatalities and serious injuries in work zones [3].

Florida continues to experience a high number of road construction work zone fatalities and serious injuries, particularly in urban counties. With a $12 billion infrastructure investment in 2022, there is increased exposure to work zones on both urban roadways and limited access roadways. Florida ranks eighth nationally for fatal work zone crashes involving commercial motor vehicles (CMVs), with 68% occurring on urban roadways and 47% on interstate highways [4].

From 2019 to 2023, Florida averaged 71 fatalities and 309 serious injuries annually in work zones. This average is consistent with the data from 2017 to 2021, indicating a persistent issue. The top ten counties with the highest work zone incidents include Orange, Broward, Duval, Hillsborough, Pasco, Miami-Dade, Seminole, Manatee, Palm Beach, and Lake [5]. There needs to be close observation of engineering solutions to reduce fatalities in work zones. Analyzing the causes in work zone crashes is necessary to identify factors leading to serious injuries and fatalities, providing insights into effective engineering solutions [6]. This paper provides a preliminary analysis of work zone crashes in four urban counties in Florida—Broward, Duval, Hillsborough, and Orange—representing significant portions of the state's work zone incidents.

**LITERATURE REVIEW**

Roadway construction work zones are critical areas where safety concerns are paramount for both workers and motorists. The increasing number of infrastructure projects and the inherent risks associated with these zones necessitate a thorough examination of safety measures and analysis techniques. This literature review synthesizes recent research findings, providing a comprehensive overview of the current state of knowledge on safety analysis in roadway construction work zones.

Managing safety in roadway construction work zones involves multiple challenges. Temporary measures and safety protocols are essential for protecting construction workers within these zones. The presence of heavy machinery, the need for continuous movement of vehicles, and the temporary nature of safety installations contribute to the complex safety dynamics. For instance, the Federal Highway Administration (FHWA) reported 669 fatalities in work zone crashes in 2014, averaging about 1.8 work zone fatalities per day. These statistics underscore the need for effective safety measures to mitigate these risks [7]. Recent studies highlight the causal relationship between work zones and crash occurrences, emphasizing the need for advanced risk analysis techniques for safety evaluation. The frequent changes in traffic patterns and the proximity of workers to moving traffic create an environment where accidents are more likely to occur, making it imperative to develop robust safety strategies [8].

Driver behavior is a significant factor contributing to work zone accidents. Studies indicate that driver error is responsible for a majority of crashes in these areas. Research found that only 6% of drivers exhibit risky driving behavior, yet these drivers cause 65% of work zone crashes. Risky behaviors include aggressive lane changing, speeding, and ignoring traffic control signs, which are prevalent in work zones [9]. The use of cellular data to assess freeway work zone safety represents a significant technological advancement. Improved data collection methods allow for more accurate monitoring of work zone conditions and the identification of potential hazards. These technological tools enhance the ability of safety managers to respond promptly to emerging threats and ensure continuous protection for workers and motorists [10]. Moreover, integrating narrative text from insurance claims into safety analysis offers a unique perspective on the causes and consequences of work zone crashes. By examining detailed descriptions of accidents, researchers can identify common factors and develop strategies to address specific risks associated with work zones [11].

Machine learning methods have been increasingly employed to analyze construction work zone crashes. A study compared the performance of different machine learning techniques, offering valuable insights into their application in safety management plans. These methods enable the identification of high-risk scenarios and the development of targeted interventions to mitigate risks. Controlling driving speed in work zones is another critical aspect of safety management. Research underscores the importance of speed control for ensuring the safety of both vehicle occupants and workers. Speeding through work zones not only increases the likelihood of crashes but also exacerbates the severity of injuries sustained [12]. Meta-analyses conducted on the effect of work zones on crash occurrence reveal that the presence of road work significantly increases the odds of accidents. This finding is supported by studies that examine the temporal factors influencing crash severity in worker-involved incidents. These studies provide a nuanced understanding of the variables that contribute to higher crash rates and more severe outcomes in work zones [13, 14].

Technological advancements have played a crucial role in enhancing work zone safety. The use of cellular data allows for more accurate monitoring of work zone conditions and the identification of potential hazards. Additionally, integrating data mining and multi-criteria decision-making (MCDM) methods provides a comprehensive approach to analyzing driver behavior and improving work zone safety. A notable study utilized k-mean clustering and the Vlsekriterijumska Optimizacija I Kompromisno Resenje (VIKOR) method to analyze driving behaviors and identify risky drivers in work zones. This integrated approach offers valuable insights for transportation engineers and decision-makers [15]. Another study highlighted the significant increase in crash rates and costs associated with work zone incidents, underscoring the need for interventions to mitigate traffic hazards [16].

Case studies provide practical insights into the effectiveness of various safety measures in work zones. For instance, a driving simulator study evaluated driver reactions to different lane-shift sign configurations. Their findings suggest that specific configurations can significantly impact driver behavior and safety in work zones. The practical application of these findings can inform the design and implementation of more effective safety measures in construction work zones [17]. Other studies have examined the impact of work zone design on driver behavior and crash risk, recommending gradual lane shifts, clear signage, and adequate lane width to enhance safety and reduce driver confusion [18].

The reviewed literature highlights the multifaceted nature of safety analysis in roadway construction work zones. From advanced machine learning techniques to the utilization of cellular data and narrative text analysis, various methodologies contribute to a deeper understanding of work zone safety. As infrastructure projects continue to grow, the importance of implementing effective safety measures and continuously improving analysis techniques cannot be overstated. Future research should focus on integrating these diverse approaches to develop comprehensive safety management plans that can adapt to the dynamic conditions of roadway construction work zones.

**DESCRIPTIVE STATISTICS OF THE WORK ZONE CRASHES IN FLORIDA**

The work zone crashes analyzed span from 2016 to 2023, focusing on the selected counties previously mentioned. This phase of the study did not include human factors such as speed, driver conditions (e.g., driving under the influence), distracted driving, or demographics. Instead, the preliminary analysis evaluated the following attributes to provide a generalized idea of the factors contributing to work zone crashes:
- Type of crash
- Work zone roadway conditions (type of work zone, crashes on work zones, and type of shoulder)
- Presence of workers and law enforcement
- Weather and lighting conditions

The descriptive statistics discussed in this section include:
- The yearly distribution of crashes in the selected zones
- The yearly distribution of crashes by severity
- The distribution of crashes by type

- The distribution of crashes by work zone type
- The distribution of crashes by light condition
- The distribution of crashes by type of shoulder
- The distribution of crashes by weather conditions
- The distribution of crashes with workers present
- The distribution of crashes with law enforcement present

**Yearly Distribution of Crashes in the Selected Zones**
As previously mentioned, work zone crashes in the selected regions represent about one-third of the total work zone crashes in the state. The average yearly crashes in the four zones during the analysis period was 1,164. Figure 1 below shows the yearly distribution of the work zone crashes.

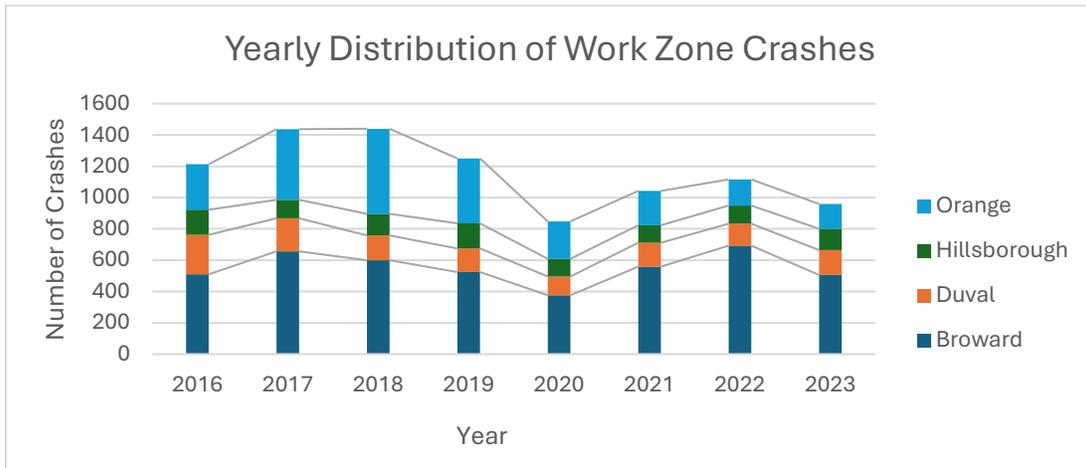

**Figure 1: Yearly distribution of work zone crashes for the selected zones.**

**Yearly Distribution of Crashes by Severity**
The severity of crashes was analyzed, with the majority being injury and property damage (non-injury) crashes. During the analysis period, there were a total of 36 fatalities, averaging 4.5 per year, and a total of 151 serious injuries, averaging 18.9 per year. Figure 2 below represents the distribution of crashes by severity for the analysis period.

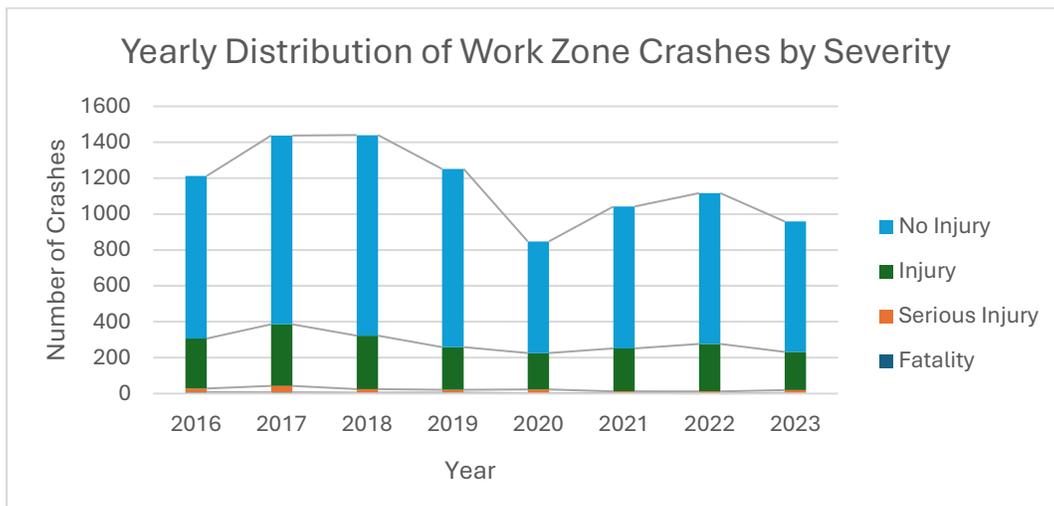

**Figure 2: Yearly distribution of work zone crashes by severity**

**Types of crashes**
An observation of the type of crashes per year showed the majority of the work zone crashes were sideswipes, rear end, other and off-road crashes. Figure 3 represents the distribution of crashes by type for the crash types with the higher frequency of crashes. In addition, a comparison of the types of crashes were done for the zones.

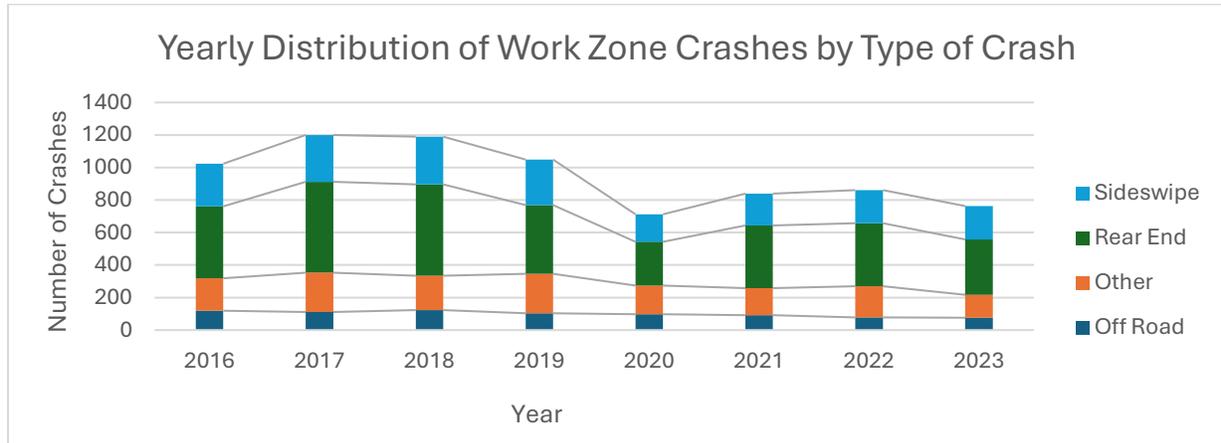

**Figure 3: Yearly distribution of work zone crashes crash type**

Further analysis showed that most fatalities and serious injuries occurred with pedestrian crashes, off-road crashes, rear-end, and left-turn crashes. Pedestrian crashes accounted for 22% of the fatalities, and off-road crashes for 19%. Rear-end crashes accounted for 26% of the serious injuries, and off-road crashes for 15%. A high percentage of crashes with fatalities and serious injuries were reported as "other," representing 31% and 17%, respectively.

**Type of Work Zone, Crashes on Work Zones, and Type of Shoulder**
The roadway conditions considered included the type of work zone, crashes in or near the work zone (labeled crashes on work zone), and type of shoulder. The types of work zones reported were intermittent or moving work, lane closure, lane shift/crossover, work on shoulder or median, and other. Most crashes occurred during lane closures, lane shifts/crossovers, and work on shoulders or medians, contributing to an average of 28%, 11%, and 28% of the crashes each year, respectively. The distribution of severe crashes indicated that work on shoulder or median contributed to 56% of fatalities, while lane closure contributed 31%. These types also contributed to 34% and 25% of serious injuries. It is noteworthy that 8% and 19% of the fatalities and serious injuries did not have information on the type of work zone.

Crashes were also classified by whether they occurred within or near the work zone. The classifications included the activity area, advance warning area, before the first work zone warning sign, termination area, and transition area. An average of 46% of crashes occurred in the activity area, followed by 17% in the transition area. Twenty-five percent of the crashes were not classified for this factor. Regarding severity, 81% of fatalities and 62% of serious injuries occurred in the activity area. Eight percent of the fatalities and 19% of serious injuries did not have information on the crash location within the work zone. Types of shoulders included paved, curb, and unpaved. Over the analysis period, 50% of crashes occurred on paved shoulders, while 36% and 14% occurred on curb and unpaved shoulders, respectively. 56% of fatalities and 55% of serious injuries occurred on paved shoulders, 25% and 28% on curb shoulders, and 19% and 17% on unpaved shoulders.

**Presence of Workers and Law Enforcement at the Work Zone**
The distribution of crashes with the presence of workers and law enforcement was observed to see if there was any improvement in safety. The data showed that more crashes occurred when law enforcement was not present. Of the crashes with law enforcement present, there were very few fatalities and serious injuries. Eighty-three percent of fatalities and 72% of serious injuries occurred when no law enforcement was present. The data also showed that an average of 36% of crashes during the analysis period occurred when workers were present. Comparing the presence of workers with the severity of crashes, there were more fatalities and serious injuries when no workers were present, contributing to about 56% and 47%, respectively. Over 28% of the overall crashes during the analysis period did not have information about the presence of workers at the work zone. Additionally, 11% of fatalities and 19% of serious injuries did not have information on the presence of workers.

**Weather and Light Conditions**
Weather conditions included blowing sand, soil, dirt; clear; cloudy; fog, smog, smoke; other; rain; and severe crosswinds. The majority of crashes occurred in clear weather, contributing to an average of 77% of the crashes. Eighty-three percent of fatalities and 76% of serious injuries occurred in clear weather, while only 3% and 7% occurred during rain. Light conditions included dark - lighted, dark - not lighted, dark - unknown lighting, dawn, daylight, dusk, other, and unknown. Most crashes occurred in daylight, followed by dark-lighted conditions, contributing to 66% and 22% of crashes, respectively. Fifty-three percent of fatalities occurred in dark-lighted conditions, and 22% occurred in daylight. For serious injury crashes, 54% occurred in daylight and 34% in dark-lighted conditions.

**Multilogit Model**
The multilogit model analysis was conducted to predict the severity of work zone crashes. This model used the severity of the crash (serious injury, injury, and non-injury) as the predictor variable, with fatality as the reference value. The dependent variables included crash type, light condition, weather condition, type of shoulder, crash in work zone, type of work zone, workers present, and law enforcement present. The results in the appendix show the full results of the model and the coefficients associated with each severity level relative to fatalities.

**DISCUSSION OF THE RESULTS**
**Crash Type**
The multilogit model analysis revealed varying impacts of different crash types on the likelihood of serious injury, injury, and non-injury outcomes. Positive coefficients for bicycle, rear-end, and sideswipe crashes indicate a higher likelihood of resulting in serious injury compared to fatalities. Conversely, negative coefficients for other crash types (animal, head-on, left turn, off-road, pedestrian, right turn, and rollover) suggest a lower likelihood of serious injury relative to fatalities. For non-injury crashes, positive coefficients were found for animal, bicycle, left turn, rear-end, right turn, and sideswipe crashes, whereas the remaining crash types showed negative coefficients. Similarly, for the injury category, positive coefficients were observed for animal, bicycle, left turn, rear-end, right turn, and sideswipe crashes, with negative coefficients for other crash types. Table 1 summarizes the coefficient results for various crash type attributes.

**Table 1: Multilogit Coefficients for the Crash Type Variable Attributes**

| Crash Predictor | Outcome |
|---|---|
| Crash Attribute | Injury |
| Animal | 2.895 |
| Bicycle | 2.719 |
| Head On | -1.704 |
| Left Turn | 0.241 |
| Off Road | -0.837 |
| Other | -1.482 |
| Pedestrian | -2.587 |
| Rear End | 0.700 |
| Right Turn | 3.977 |
| Rollover | -1.848 |
| Sideswipe | 6.292 |

**Type of Work Zone, Crashes on Work Zones, and Type of Shoulder**

The analysis of work zone type, crashes on work zones, and type of shoulder attributes indicated significant variations in crash outcomes. Positive coefficients for advance warning area, termination area, intermittent or moving work, lane shift/crossover, curb, paved shoulder, and unpaved shoulder suggest a higher likelihood of serious injury compared to fatalities. Negative coefficients for other attributes indicate a lower likelihood of serious injury. For non-injury crashes, similar trends were observed, except for the curb shoulder, paved shoulder, and unpaved shoulder, which had negative coefficients. Table 2 provides a summary of the coefficients for these attributes.

**Table 2: Multilogit Coefficients: Type of Work Zone, Crashes on Work Zones, and Type of Shoulder**

| Crash Predictors | Outcome |
|---|---|
| Crash Attribute | Injury |
| Advance Warning Area | 4.773 |
| Before the First Work Zone Warning Sign | -1.300 |
| Termination Area | 3.318 |
| Transition Area | -1.239 |
| Intermittent or Moving Work | 4.525 |
| Lane Closure | -1.975 |
| Lane Shift/Crossover | 4.936 |
| Other | -1.705 |
| Work on Shoulder or Median | -2.213 |
| Curb Shoulder | 5.170 |
| Paved Shoulder | 4.850 |
| Unpaved Shoulder | 4.637 |

**Presence of Workers and Law Enforcement at the Work Zone**

The presence of workers and law enforcement significantly impacts crash outcomes. Positive coefficients for all severity categories indicate a lower likelihood of fatalities when workers or law enforcement are present. Higher coefficients for serious injuries suggest that their presence effectively reduces fatality risks. Table 3 summarizes the coefficients for these attributes.

**Table 3: Multilogit Coefficients: Presence of Workers and Law Enforcement at the Work Zone**

| Crash Predictors | Outcome |
|---|---|
| Crash Attribute | Injury |
| Workers Present (N) | 0.590 |
| Workers Present (Y) | 0.993 |
| Law Enforcement Present (N) | 1.476 |
| Law Enforcement Present (Y) | 2.092 |

**Weather and Light Conditions**

The analysis of weather and light conditions revealed varying impacts on crash severity. Positive coefficients for dawn, daylight, fog, smog, smoke, and rain indicate a higher likelihood of serious injury compared to fatalities. Negative coefficients for other conditions suggest a lower likelihood of serious injury. Table 4 provides a summary of these attributes.

**Table 4: Multilogit Coefficients for Weather and Light Conditions**

| Crash Predictors | Outcome |
|---|---|
| Crash Attribute | Injury |
| Dark - Not Lighted | -0.244 |
| Dark - Unknown | 6.335 |
| Dawn | 7.485 |
| Daylight | 1.728 |
| Dusk | 0.728 |
| Other | -2.716 |
| Unknown | -5.292 |
| Clear | 1.802 |
| Cloudy | 1.934 |
| Fog, Smog, Smoke | 8.358 |
| Other | -5.603 |
| Rain | 3.084 |
| Severe Crosswinds | -2.188 |

The results from Tables 1 to 4 highlight the importance of various attributes in predicting crash severity. The analysis indicates that 55% of serious injury, 38% of injury, and 42% of non-injury categories have negative coefficients, presenting a high likelihood of resulting in a fatality. However, 35% of the attributes showed positive coefficients for all severity levels, indicating a lower likelihood of fatalities for those attributes.

These attributes are represented in each of the crash categories in this study, i.e.

- For the crash in work zone: advance warning area, termination area,
- For the crash type: bicycle crashes, rear end crashes, sideswipe,
- For presence of law enforcement and workers: Yes and No
- For the light and weather condition: dawn, daylight, intermittent or moving work, lane shift/crossover, fog, smog, smoke weather conditions

Equation (i) shows the log odds equation.

$$log\left(\frac{P(category)}{P(reference)}\right) = \beta_{0,category} + \beta_{attribute\ 1,category} \cdot Attribute1 + \beta_{attribute\ 2,category} \cdot Attribute\ 2 + \cdots \quad (i)$$

For each equation representing a category

- The reference category is the fatalities
- The intercept term ($\beta_{0,category}$ which are the $\beta_{0,serius\ injury}$ $\beta_{0,injury}$ and $\beta_{0,non-injury}$) represents the log odds of the outcome when all predictors are zero.
- Each coefficient (e.g., $\beta_{attribute\ 1}$, $\beta_{attribute\ 2}$... which are $\beta_{animal}, \beta_{bicycle}, \beta_{head-on}$ etc) represents the change in the log odds of the outcome for a one-unit change in the predictor variable.
- Positive coefficients indicate an increase in the log odds of the outcome relative to the reference category, while negative coefficients indicate a decrease.

For the serious injury the log odds equation are as shown in equation (ii), (iii) and (iv)

$$log\left(\frac{P(injury)}{P(fatalities)}\right) = -1.841 + 2.894.Animal + 2.719.Bicycle - 1.704.Head-on \ldots \text{(ii)}$$

$$log\left(\frac{P(serious\ injury)}{P(fatalities)}\right) = 1.330 - 5.544.Animal + 3.817.Bicycle - 0.706.Head\ on \ldots \text{(iii)}$$

$$log\left(\frac{P(no\ injury)}{P(fatalities)}\right) = 12.319 + 3.637.Animal + 1.186.Bicycle - 1.175.Head-on \ldots \text{(iv)}$$

**CONCLUSION AND FUTURE WORKS**

The findings from this study using the multilogit model provide significant insights into the factors contributing to the severity of crashes in construction work zones in Florida. By examining attributes such as crash type, work zone conditions, presence of workers and law enforcement, weather, and light conditions, the analysis highlights critical areas where improvements can be made to enhance work zone safety. One key finding is that bicycle, rear-end, and sideswipe crashes are more likely to result in serious injuries compared to fatalities. This indicates a need for targeted interventions to address these specific types of crashes. On the other hand, crash types such as animal, head-on, and pedestrian crashes were found to have a lower likelihood of resulting in serious injuries compared to fatalities. This differentiation suggests that safety measures can be more precisely tailored to the nature of the crash type.

The analysis also revealed that certain work zone conditions, such as the advance warning area, termination area, and lane shift/crossover, are associated with a higher likelihood of serious injuries. Enhancing safety measures in these specific work zone areas could significantly reduce the severity of crashes. For instance, improving signage and alerts in these critical sections could help drivers navigate more safely through work zones. The presence of workers and law enforcement emerged as a significant factor in reducing the likelihood of fatalities. Crashes were notably more severe when no law enforcement was present, underscoring the importance of increasing law enforcement presence in work zones to deter risky driving behaviors and enhance overall safety. Weather and light conditions also play a crucial role in the severity of crashes. The majority of crashes occurred in clear weather and daylight, but serious injuries were more likely during dawn, fog, smog, and smoke conditions. This finding suggests a need for improved safety measures tailored to this specific weather and light conditions, such as additional lighting or visibility aids. These insights suggest several actionable steps for improving work zone safety. Implementing advanced warning systems and clear signage in critical work zone areas can help alert drivers and reduce crash severity. Increasing the presence of law enforcement in work zones can help deter risky driving behaviors and enhance overall safety. Utilizing technologies like DeepSORT for real-time monitoring and tracking can provide continuous safety assessments and immediate alerts for potential hazards. Developing weather-specific safety measures can help mitigate the increased risks associated with certain conditions.

Future research should aim to expand the model to include more detailed human factors, such as driver behavior and traffic conditions, to provide a more comprehensive understanding of crash dynamics in work zones. Additionally, integrating machine learning algorithms with real-time data collection can offer more robust predictive capabilities, enabling proactive safety management. By extending the geographic scope of the analysis to include more counties and different types of roadways, a broader understanding of work zone safety across diverse environments can be achieved. In conclusion, this study underscores the potential of using advanced modeling and technological tools to enhance safety in construction work zones. Addressing the identified risk factors and implementing targeted interventions can significantly improve safety outcomes for both workers and motorists in these high-risk environments. Developing comprehensive safety protocols that integrate these findings with ongoing technological advancements will be crucial in achieving the goal of reducing fatalities and serious injuries in construction work zones.

**AUTHOR CONTRIBUTIONS**
The authors confirm their contributions to the paper as follows: study conception and design: Doreen Kobelo, Mohamed Khalafalla; data collection: Tatiana Deslouches; analysis and interpretation of results: Tatiana Deslouches, Tejal Mulay; draft manuscript preparation: Tatiana Deslouches, Tejal Mulay, Doreen Kobelo, Mohamed Khalafalla. All authors reviewed the results and approved the final version of the manuscript.